\documentstyle[epsf,sprocl]{article}

\input{psfig.sty}

\def\be{\begin{equation}}
\def\ee{\end{equation}}
\def\bea{\begin{eqnarray}}
\def\eea{\end{eqnarray}}
\def\Dslash{D{\hskip -0.22cm}\slash} 

\begin{document}

\title{A Theoretical Review of Axion\footnote{Talk presented at cosmo-99,
ICTP, Trieste, Italy, 28 September 1999.}}

\author{Jihn E. Kim}

\address{Department of Physics and Center for Theoretical Physics \\
Seoul National University, Seoul 151-742, Korea } 

\maketitle\abstracts{It is emphasized that the existence of a very light 
axion is consistent with the strong CP invariance and cosmological
and astrophysical constraints.  The attempt to embed the very light axion
in superstring models is discussed.}

\section{The Strong CP Problem}

The standard model $SU(3)\times SU(2)\times U(1)$ describes the
weak and electromagnetic interactions very successfully. The strong
interaction part, quantum chromodynamics, is proven to be successful
at perturbative level, but the study of nonperturbative effects are not 
so successful. Because of the lack of a calculational tool of the
nonperturbative effects, the frequently used method for the study
of QCD at low energy is the symmetry principle. At energy scales below
the confinement and chiral symmetry breaking scale, some symmetries
of QCD are manifest in strong interaction dynamics. Baryon number is
known to be conserved. Chiral symmetry is broken. 

Discrete symmetries are believed to be conserved in the process of
confinement. Here we are interested in CP. If QCD conserves CP,
the CP symmetry will be preserved in strong interactions at low energy.
If QCD violates CP, its effect will be shown in low energy strong
interaction dynamics. QCD before 1975 was described by
\begin{equation}
{\cal L}= -\frac{1}{2g^2}{\rm Tr\ }F_{\mu\nu}F^{\mu\nu}
+\bar q(i\Dslash-M_q)q
\end{equation} 
where $q$ and $M_q$ are quark and quark mass matricies, respectively.
After the discovery of the instanton solution in non-Abelian gauge
theories,\cite{bpst} it is known that the $\theta$-term must be
considered,\cite{theta}
\begin{equation}
{\cal L}_\theta=\frac{\theta}{16\pi^2}{\rm Tr}\ F_{\mu\nu}\tilde F^{\mu\nu}.
\end{equation}
Note that ${\cal L}_\theta$ is odd uner P or T discrete transformation.
Therefore, this term violates the CP invariance. Thus QCD contains a
term violating CP symmetry. If QCD violates the CP symmetry, it must be
revealed in strong interaction dynamics. For example, we may expect a 
CP violating static property of neutron, the electric dipole moment
of neutron $d_n$. The experimental upper limit of $d_n$ is known to
be $0.63\times 10^{-25}\ e\cdot$cm.\cite{dn} On the other hand, if
strong interaction violates the CP invariance at the full strength,
then we expect $d_n\sim 10^{-14}\ e\cdot$cm. Therefore, the vacuum
angle is restricted to a tiny region
\begin{equation}
|\theta|< 10^{-9}.
\end{equation}
The question, $\lq\lq$Why is the vacuum angle so small", is the
strong CP problem. If we treat $\theta$ as O(1) parameter, then the
above observation excludes QCD as the theory of strong interactions.
In this case, different $\theta$'s describe different universes. Since 
QCD is known to be so successful except for the strong CP problem, it is
better to keep QCD as the theory of strong interactions. It is
desirable to resolve the strong CP problem with QCD untouched.

There have been several attempts toward the solution of the strong
CP problem, one even incorrectly asserting that there is no
strong CP problem.\cite{marshak}\footnote{This reference~[4]
starts with an assumption on CPT invariant $|n\rangle$ vacua and obtains
the CPT odd $|\theta\rangle$ vacuum. But in Yang-Mills theories,
$|n\rangle$ vacua are not CPT invariant, and the $|\theta\rangle$
vacuum is CPT invariant.} 

One class of solutions employs CP
symmetry at the Lagrangian level, and require the induced vacuum 
angle (in the process of introducing weak CP violation) sufficiently
small.\cite{natural} These are called natural solutions.
Here, the weak CP violation is through spontaneous symmetry
breaking or soft breaking. Namely, the Kobayashi-Maskawa type
weak CP violation is not possible except for the Nelson-Barr type.
The reason for a possible Kobayashi-Maskawa weak CP violation in
the Nelson-Barr type strong CP solution is that the spontaneous CP
violation here is introduced at a super high energy scale and
hence at the electroweak scale CP is already violated as in the
Kobayashi-Maskawa model. In this class of models, at tree level
$\theta=0$, which implies $Arg\ Det\ M_q=0$. This is attained by
assuming a CP invariant Lagrangian including the $\theta$ term
and specific symmetries. Possible symmetries used for this purpose
are: left-right symmetry, $U(1)$ gauge symmetry, permutation symmetry,
and other discrete symmetries.

The other solutions are the $m_u=0$ solution and the axion solution.

Here, the most attractive solution is the axion solution, making
$\theta$ a dynamical variable. A dynamical $\theta$ is equivalent to
a pseudoscalar field which we call axion. If the axion has a potential,
then in the evolving universe the minimum of the vacuum will be
chosen. The axion solution guarantees that $\theta=0$ is the
minimum of the $\theta$ potential, which is discussed below.

\section{The Peccei-Quinn Solution}

The axion solution is a dynamical solution. Peccei and Quinn~\cite{pq}
showed that $\theta=0$ at the point $dV/d\theta=0$, which has been
shown later by Vafa and Witten,\cite{vw}
\begin{equation}
{\cal L}=-\frac{1}{4}F^2+\bar q(i\Dslash-M_q)q+\theta\frac{g^2}{32\pi^2}
F\tilde F
\end{equation}
where we suppressed the Greek indices for space-time in $F$. Let us treat
$\theta$ as a coupling. After integrating the quark fields out, the
generating functional in the Euclidian space is given by
\begin{equation}
\int [dA_\mu]\prod_i{\rm Det}(\Dslash+m_i)
\exp{\{-\int d^4x[\frac{1}{4g^2}F^2-i\theta\frac{1}{32\pi^2}F\tilde F]\}}.
\end{equation}
Note that the resulting functional has a specific form of the $\theta$
dependence. In the Euclidian space, corresponding to the eigenstate
$\psi$ of $i\Dslash$, $i\Dslash\psi=\lambda\psi$, there corresponds to
the other eigenstate of $i\Dslash$, $i\Dslash(\gamma_5\psi)=-\lambda
(\gamma_5\psi)$. Thus, the nonzero real eigenvalues of $i\Dslash$ are
paired with opposite sign and the identical magnitude. For $N_0$ number
of zero modes, we can show that
\begin{equation}
{\rm Det\ }(\Dslash+m_i)=\prod_i (-i\lambda+m_i)=m_i^{N_0}(m_i^2+
\lambda^2)>0.
\end{equation}
Thus, the generating functional is bounded by usung the
Schwarz inequality in view of Eq.~(6),
\begin{eqnarray}
&\exp[-\int d^4x V[\theta]]\equiv |\int[dA_\mu]\prod {\rm Det\ }
(\Dslash+m_i)\exp(-\int d^4x{\cal L})|\nonumber\\
&\le \int [dA_\mu]|\prod {\rm Det}(\Dslash+m_i)\exp(-\int d^4x{\cal L})
|\nonumber\\
&=|\int[dA_\mu]\prod{\rm Det\ }(\Dslash+m_i)\exp[-\int d^4x
{\cal L}(\theta=0)] |\\
&=\exp(-\int d^4x V[0])\nonumber
\end{eqnarray}
where ${\cal L}=(1/4g^2)F^2-i\theta\{F\tilde F\}$, and the simplified
notation $\{\ \ \}$ includes a factor $1/32\pi^2$. The above
inequality guarantees
\begin{equation}
V[\theta]\ge V[0].
\end{equation}
Instanton solutions have integer values for $\int d^4x\{F\tilde F\}$,
hence $V[\theta]$ is periodic with the $\theta$ period of $2\pi$,
$\theta\rightarrow \theta+2n\pi$. The above agument is for a nonzero
up quark mass. If $m_u=0$, $\theta$ is unphysical and there is no
strong CP problem.

As a coupling, any $\theta$ defines a good theory (or universe). The 
shape of $V$ as a function of $\theta$ is

\begin{figure}
\epsfxsize=60mm
\centerline{\epsfbox{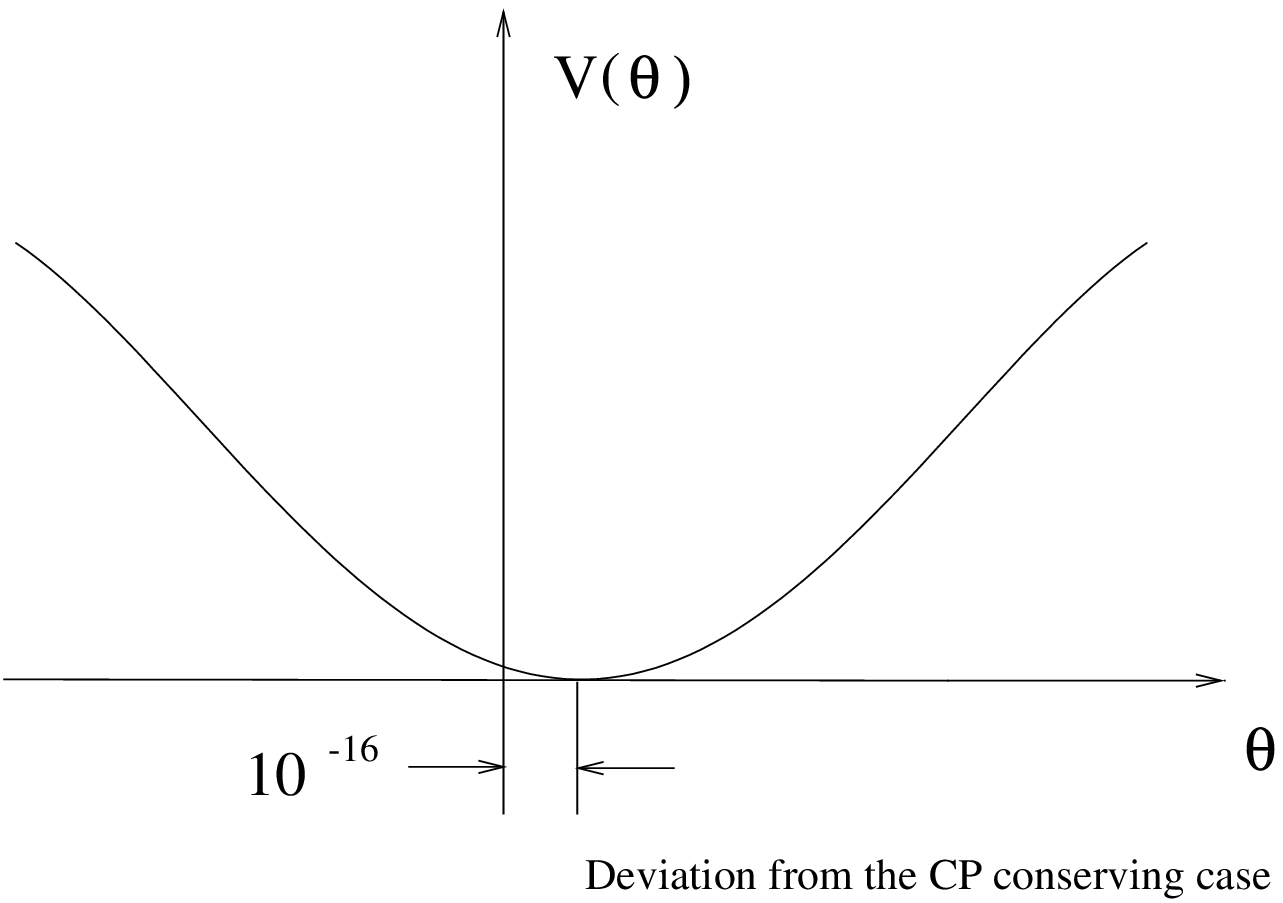}}
\end{figure}
\centerline{Fig.~1.
{\it Shape of $V[\theta]$. The KM weak CP introduces}~\cite{gr} 
$|\theta| \simeq 10^{-16}$.}
\vskip 0.3cm
Here, the axion solution of the strong CP problem is transparent. If
we identify $\theta$ as a pseudoscalar field,
\begin{equation}
\theta=\frac{a}{F_a}
\end{equation}
the vacuum angle is chosen at $\theta\simeq 0$, realizing the
almost CP invariant QCD vacuum. Note the key ingredients of this
axion solution. Firstly, the theory introduces an axion coupling
$a\{F\tilde F\}$. Second, there is no axion potential except
that coming from the $F\tilde F$ term, otherwise our proof does not
go through. Third, there necessarily appears a mass parameter $F_a$,
the so-called axion decay constant. One possible example for 
the axion is to introduce a
Goldstone boson, using a global $U(1)$ symmetry.~\cite{pq} As explained
above Peccei-Quinn showed that $\theta=0$ is the minimum of the
potential, which is meaningful only if there is a dynamical field
$a$. Later, Weinberg and Wilczek explicitly showed that the model
contains the axion.~\cite{ww} It was immediately known that the
PQWW axion does not exist,\cite{peccei} and hence there soon 
appeared a flurry of natural solutions.~\cite{natural} 

To have the anomalous coupling, the global symmetry must have a
triangle anomaly in $U(1)_{\rm global}\times SU(3)_c\times SU(3)_c$.
This anomalous coupling introduces an axion decay constant which
can be large for the very light axion models.\cite{invisible,dfsz}
Soon after the invention of the very light axion, it has been
known that the astrophysical and cosmological bounds restrict 
the axion decay constant in the region, $10^9{\ \rm GeV}\le F_a
\le 10^{12}\ {\rm GeV}$.\cite{astro,cos}
Also, it was known that a more ambitious composite axion can be 
constructed.\cite{comp}

Another interesting possibility is that there results an axion with
the above properties from a more fundamental theory such as from
the string theory. This possibility introduces a nonrenormalizable
interaction $aF\tilde F$. Indeed, superstring models have a host
of moduli fields which do not have potentials at the compactification
scale. But some of the moduli have the desired anomalous 
couplings,\cite{witten} and becomes the axion. In this case, the 
so-called superstring axions are expected to have the decay constant 
near the Planck scale. But the exact magnitude depends on the details
of the model. 

The mass of the very light axion is an important parameter in the
evolving universe. The allowed range of the decay constant gives
tens of micro-eV axion mass,
\begin{equation} 
m_a=0.6\times 10^7\ \frac{\rm eV}{F_a^{\rm GeV}}
\end{equation}
where $F_a^{\rm GeV}=F_a/{\rm GeV}$.

The neutron electric dipole moment in the $\theta$ vacuum is
given by~\cite{quint}
\begin{equation}
\frac{d_n}{e}=O(1)\frac{m_u\sin\theta}{f_\pi^2[2Z\cos\theta+(
1+Z)^2]^{1/2}}
\end{equation}
where $Z=m_u/m_d$ is the ratio of the current quark masses.
For $m_u<2\times 10^{-13}$~GeV, the neutron electric dipole 
moment is satisfied even for O(1) $\theta$.

The very light axion physics is closely connected to the study
of the evolution of the universe. The domain wall problem~\cite{sikivie}
must be studied in specific models. Usually, inflation needed in
supergravity models with the condition on the reheating temperature
$T_R<10^9$~GeV \cite{ekn} does not lead to the axionic domain
wall problem.

Theoretically, the introduction of the Peccei-Quinn U(1) global
symmetry is ad hoc. It is better if the axion arises from a
fundamental theory. In this spirit, it is most important to
draw a very light axion from superstring theory. If we cannot,
how can we understand the strong CP problem?

\section{Embedding the Very Light Axion in Superstring}

The pseudoscalar moduli fields in D=10 superstring is $B_{MN}\
(M,N=0,\cdots,9)$ among the bosonic fields $G_{MN}, B_{MN}$ and
the dilaton. Upon compactification to D=4, $B_{\mu\nu}\ (\mu,\nu
=0,\cdots,3)$ turns out to be a pseudoscalar field. Dual 
transformation of $B_{\mu\nu}$ defines a pseudoscalar $a$ as
\begin{equation}
\partial^\sigma a\propto\epsilon^{\mu\nu\rho\sigma}H_{\mu\nu\rho};
H_{\mu\nu\rho}={\rm field\ strength\ of\ }B_{\mu\nu}\propto
\epsilon_{\mu\nu\rho\sigma}\partial^\sigma a.
\end{equation}
Of course, $B_{\mu\nu}$ does not have renormalzable couplings to 
matter fields, hence there is no potential for $a$ since the
possible derivative coupling $\partial^\mu a\psi\gamma_\mu\gamma_5
\psi$ does not lead to a potential term. If we consider a field
strenth 
$H_{\mu\nu\rho}$ to obtain couplings, it is invariant under
a shift $a\rightarrow a+c$. This consideration does not lead to
an anomalous coupling needed for an axion. 

However, the gauge invariant D=10 field strength of $B_{MN}$ is
not $H=dB$,\footnote{In this paragraph we use the differential form.}
but is~\cite{green}
\begin{equation}
H=dB+\omega^0_{3Y}-\omega^0_{3L}
\end{equation} 
where the Yang-Mills Chern-Simmons form is tr$(AF-A^3/3)$ and the
Lorentz Chern-Simmons form is tr$(\omega R-\omega^3/3)$. The 
Chern-Simmons forms satisfy $d\omega^0_{3Y}={\rm tr\ }F^2$ and
$d\omega^0_{3L}={\rm tr}\ R^2$. Therefore,
\begin{equation}
dH=-{\rm tr\ }F^2+{\rm tr\ }R^2,
\end{equation}
and the equation of motion for $a$ is
\begin{equation}
\Box a=-\frac{1}{M}[{\rm Tr\ }F_{\mu\nu}\tilde F^{\mu\nu}
-{\rm Tr\ }R_{\mu\nu}\tilde R^{\mu\nu}]
\end{equation}
which implies $aF\tilde F$ coupling which is needed for the
axion interpretation of $a$.\cite{witten} This is the so-called
model-independent axion. Here, $M$ is about the compactification
scale suppressed by a factor and corresponds to the axion decay
constant.\cite{choikim} 

To cancel the Yang-Mills anomaly, one
should introduce the Green-Schwarz term,\cite{green} $S_{GS}\sim
\int(B{\rm tr}F^4+\cdots)$. The Green-Schwarz term gives the 
needed anomalous coupling for pseudoscalars $B_{ij}\ (i,j=4,\cdots,
9)$
\begin{equation}
B_{ij}\epsilon^{\mu\nu\rho\sigma}F_{\mu\nu}F_{\rho\sigma}
\langle F_{kl}\rangle\langle F_{pq}\rangle\epsilon^{ijklpq}.
\end{equation}
Here, we have model-dependent axions $a_k\sim \epsilon_{ijk}B_{ij}$,
the number of which is the second Betti number.\cite{witten1}
Unlike the model-independent axion, the model-dependent axions receive
nonvanishing superpotential terms from the world-sheet instanton effect,
\begin{equation}
\int_{\Sigma_J} d^2z \omega^I_{ij}(\partial X^i\underline
\partial X^{\underline j}-\underline\partial X^i\partial X^{\underline
j})=2\alpha^\prime\delta_{IJ} 
\end{equation}
where $\alpha^\prime$ is the string tension, and $\omega=4\pi^2 {\rm Re}
(T_I)\omega^I$. The internal space volume is given by\cite{mth}
$V_6=(1/3!)\int \omega\land\omega\land\omega\approx 
(1/6)(4\pi^2{\rm Re}T)^3
(2\alpha^\prime)^3$. So the model-dependent
axion cannot be a candidate for the low energy QCD axion unless
the potential is sufficiently suppressed.\cite{wen}

Thus, the model-independent axion is a good candidate for
the QCD axion. However, it has two serious problems:
\\
\indent (A) The decay constant problem-- $F_a$ is too large, $\sim 
10^{16}$~GeV,\cite{choikim} and\\
\indent (B) The hidden sector problem-- We need a hidden sector confining
force for 
supersymmetry breaking around $10^{10-13}$~GeV. If so, the
dominant contribution 
to the model-independent axion comes from
the hidden sector anomaly, 
\newpage
\begin{figure}
\epsfxsize=80mm
\centerline{\epsfbox{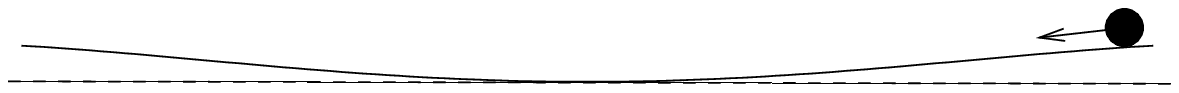}}
\end{figure}
\centerline{Fig. 2. {\it The almost flat axion potential.}}
\vskip 0.3cm
\noindent $m_a\simeq \Lambda_h^2/F_a$. Then, this 
cannot be the needed axion for the strong CP problem. With two confining
forces with scales of $\Lambda_h$ and $\Lambda_{QCD}$, the potential can
be written as
\begin{equation}
V\sim -\Lambda^4_{QCD}\cos(\theta+\alpha)-\Lambda^4_h\cos(\theta_h+\beta)
\end{equation}
where $\alpha$ and $\beta$ are constants, and $\Lambda_h\gg\Lambda_{QCD}$.
To settle both $\theta_h$ and $\theta$ dynamically at zero, we need
two axions. But as shown above, we have only one axion for this 
purpose, the model-independent axion.

Both of the above problems are difficult to circumvent.\footnote{See, 
however, Lalak et al.\cite{lalak}} 

The approximate global symmetries may be a way out from this dilemma.
\cite{shafi} Discrete symmetries may forbid sufficiently many terms so that
the Peccei-Quinn symmetry violating terms can appear only at $d\ge 9$.
One such example is $Z_N$ symmetry (e.g. $N=3$) in theories without
hidden sector quarks.\cite{gkn} 

\section{Cosmology with Axion}

In the hot cores of stellar objects, the axion production can occur
through $\gamma+e({\rm or\ }Z)\rightarrow a+
e({\rm or\ }Z)$, $n+n\rightarrow n+n+a$, $\gamma+e\rightarrow a+e$,
$e^++e^-\rightarrow a+\gamma$, etc. For a sufficiently large $F_a$,
axions produced in the stellar core can escape the star since the 
rescattering cross section is small. If its production rate is too
large $(\sim 1/F_a^2)$, it takes out too much energy from the core. 
Thus, there results the upper bounds on $F_a$ from star evolutions.
The best bound is obtained from the study of supernovae.
\cite{kang,astro,choikang} 
Thus, the solar axion search may not succeed which
needs $F_a\sim 10^7$~GeV.

The lifetime of $a$ is extremely long and hence can be treated in most cases
as a stable particle. The classical coherent states of $a$ will 
oscillate around the minimum $\langle a\rangle=0$. When can this 
happen? It is around $T_1\simeq 1$~GeV, not around the axion scale of 
$F_a$ since the axion potential is extremely flat.
Its existence is felt when the expansion rate is smaller than the
oscillation rate of the classical axion field, viz. 
$3H<m_a$.\cite{cos} For $T<T_1$, 
the classical axion field $\langle a\rangle$ begins
to roll down the hill. After this happens, the Hubble expansion is 
negligible and the $\langle a\rangle$ equation leads to a conserved
$m_aA^2$ (where $A$ is the amplitude of the classical axion field)
in the
comoving volume. 
This coherent axion field carries energy
density behaving 
like nonrelativistic particles and its
\newpage 
\begin{figure}
\epsfxsize=100mm
\centerline{\epsfbox{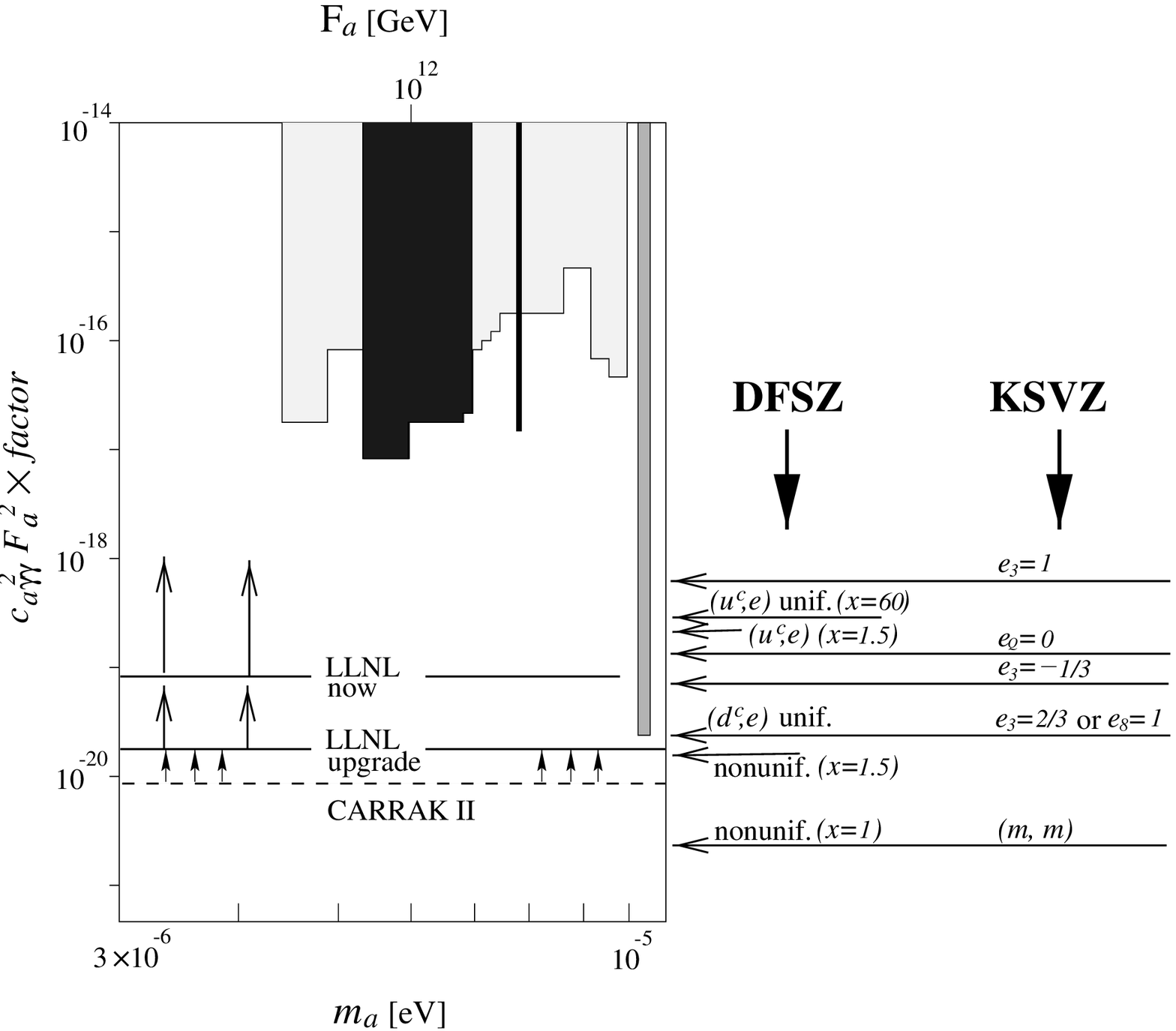}}
\end{figure}
\centerline{Fig. 3. {\it Axion search experiments. The model predictions 
are shown.}\cite{search}}
\vskip 0.3cm
\noindent contribution to cosmic energy is~\cite{review}
\begin{equation}
\Omega_ah^2\simeq 0.13\times 10^{\pm 0.4}\Lambda_{200}^{-0.7}
f(\theta_1)\left(\frac{10^{-5}{\rm eV}}{m_a}\right)^{1.18}N^2_{DW}
\end{equation}
where $\theta_1$ is the $\theta$ value at the cosmic temperature
$T_1$. These considerations restrict $F_a$ as
\begin{equation}
10^9\ {\rm GeV}\le F_a\le 10^{12}\ {\rm GeV}.
\end{equation}
In this scenario, cold axions are packed around us , for which the 
axion search experiments are performed. In this search one probes
the axion--electromagnetic coupling of $a{\bf E\cdot B}$.\cite{sik83}
The current status is exhibited in Fig.~3.

Depending on models, there can exist domain walls, but in supersymmetric
models the requirement of $T_{RH}<10^9$~GeV gives a sufficient 
dilution of the dangerous domain walls. 
Also there can exist hot axions produced by
vibrations of axionic strings when they are formed around $T\sim 
F_a$.\cite{shellard} These hot axions are also diluted by inflation
with $T_{RH}<10^9$~GeV.

\vskip 0.2cm
\noindent {\bf Quintessence idea}
\vskip 0.2cm
Recently, there is an evidence that the cosmological constant
is very tiny,\cite{perl} which is another difficult problem for the
cosmological constant. Since the axion potential is almost flat, there
may be a mechanism to have a very small cosmological constant within
the axion idea. We start with the assumption that at the minimum
of the potential the cosmological constant is zero.

We note that the massless quark solution of the strong CP problem 
leads to a flat $\theta_h$ direction. Eventually, we will identify
this $\theta_h$ direction as the hidden sector(h-sector) axion 
direction. If we break the global symmetry by a tiny h-sector quark
mass, the degeneracy is broken feebly. A random value of $\theta_h$
will give a generic value of the potential determined by the
nonvanishing h-quark mass. This generic value of the potential 
energy is expected to be $(0.003\ {\rm eV})^4$ so that it explains
the Type 1a data.\cite{perl} If $\theta_h$ is a coupling, then the
random value of $\theta_h$ gives a true cosmological constant, i.e.
it is zero. If $\theta_h$ is a dynamical field such as an axion, 
then the cosmological constant is nonzero like in axion models.
We will call this dynamical $\theta_h$ with a currently interesting
cosmological constant a {\it quintessence}. The axion quintessence
needs a potential height of order $10^{-47}$~GeV$^4$ and 
$F_a\sim M_P$, i.e. $m_a\sim 10^{-33}$~eV, for it to dominate the
mass density of the universe recently. In terms of the known scales,
$M_P=2.44\times 10^{18}$~GeV and $v\simeq 247$~GeV, we obtain
a small energy density $v^{n+4}/M_P^n$. For $n=3$, 
\begin{equation}
\frac{v^7}{M_P^3}\sim 4\times 10^{-39}\ {\rm GeV}^4
\end{equation}
which can be a reasonable candidate for the vacuum energy with
a further suppression by coupling constants. How can one forbid
$n=1,2$ but allow $n=3$ in supergravity?

With a (almost) massless h-quark, the h-sector instanton potential
is almost flat. Then the axion corresponding to the h-sector
can be a quintessence.\cite{kim} For this idea to work, we must 
introduce at least one model-dependent axion so that two axions
survive. If two axions are present with $F_1$ and $F_2$ and two
explicit scales $\Lambda_1$ and $\Lambda_2$ break the
symmetries, the larger $F$ corresponds to the smaller $\Lambda$.
Because the h-sector instanton potential is made almost
flat by almost massless h-quark,\cite{kim} the smaller symmetry
breaking scale $(v^7/M_P^3)^{1/4}$ corresponds to the larger $F$,
i.e. the Planck scale decay constant. This solves the axion decay
constant problem by lowering the decay constant of the QCD axion to
$10^{12}$~GeV. 

Now the problem is how to save a model-dependent axion. For this
we have to assume that many singlets do not develop vacuum
expectation values.\cite{kim}

\section{Conclusion}

We have shown that:

\noindent
i) The strong CP problem is a serious problem.\\
ii) But there are solutions, natural~\cite{natural} and automatic.
\cite{pq,ww,invisible,dfsz}\\
iii) The very light axion solution is the most attractive one.
\cite{invisible,dfsz} Here, the\\
\indent\hskip -0.3cm weak CP violation is of 
the Kobayashi-Maskawa type.\\
iv) The very light axion can close the universe,\cite{cos}
and can be detected.\\
v) Superstring models have two problems housing the very light
axion. One\\
\indent\hskip -0.3cm possible scenario with quintessence is also discussed.

\section*{Acknowledgments}
This work is supported in part by the Korea Research Foundation,
Korea Science and Engineering Foundation, and the BK21 program
of the Ministry of Education.

\end{document}